# Visible astro-comb filtered by a passively-stabilized Fabry-Perot cavity

Yuxuan Ma,[1] Fei Meng,[1,3] Yizhou Liu,[1] Fei Zhao,[2] Gang Zhao[2], Aimin Wang,[1] and Zhigang Zhang[1,a)]

[1]State Key Laboratory of Advanced Optical Communication System and Networks, School of Electronics Engineering and Computer Science, Peking University, Beijing 100871, China
[2]National Astronomical Observatories of China, Chinese Academy of Science, Beijing100012, China
[3]National Institute of Metrology, Beijing 100029, China

We demonstrate a compact 29.3 GHz visible astro-comb covering the spectrum from 560nm to 700nm. A 837 MHz Yb:fiber laser frequency comb phase locked to a Rb clock served as the seed comb to ensure the frequency stability and high side mode suppression ratio. After the visible super-continuum generation, a cavity-length-fixed Fabry-Perot cavity made by ultra-low expansion glass was utilized to filter the comb teeth for eliminating the rapid active dithering. The mirrors were home-made complementary chirped mirrors pair with zero dispersion and high reflection to guarantee no mode skipping. These filtered comb teeth were clearly resolved in an astronomical spectrograph of 49,000 resolution, exhibiting sharp linetype, zero noise floor, and uniform exposure amplitude.

## INTRODUCTION

Over the past decade, optical frequency combs (OFCs) have been proposed to be an ideal wavelength calibrator of astronomical spectrographs ('astro-comb') [1, 2]. The astro-combs can supply tens of thousands equally spaced and individually (although not fully) resolved lines to astronomical spectrographs, therefore the resolution of radial velocity (RV) measurement can be significantly promoted from m/s level to cm/s level. This improvement paves the way for many astrophysical applications, such as the mass measurement of Earth-like exoplanets.

There have been number of researches on astro-combs reported and quite a few institutions involved [3-9]. Since the direct generation of broadband and >10 GHz comb is difficult, conventional astro-combs generally utilize Fabry-Perot cavities (FPCs) to multiply the frequency spacing to meet the spectrograph resolution [3-6]. Recently, some novel astro-combs without FPC are reported. They directly generate several tens of GHz mode spacing based on the electro-optic (EO) comb [7] or microresonator combs [8, 9]. However, these astro-combs can only cover the infrared wavelength. For the calibration of Sun-like stars which focus on the visible region, FPCs are still needed by the visible astro-combs.

There are many issues existing in the conventional actively locked FPCs. As is well known, the slope of error signal in Pound-Drever-Hall method depends on the linewidth of the FPC [10]. Different from the high finesse FPCs, the FPCs used in astro-combs generally have ~100 MHz level linewidth, because they need to compromise the high reflectivity and the low phase error (or dispersion) over hundred nm broadband. Therefore, the slope of error signal will be very low and thus introduces a big dithering after locking. The dithering allows the side-modes to have time-varying transmission and the CCD accumulates all of them during exposure procedure. Hence, the comb lines become "fat" and make the data fitting harder. Moreover, the scheme of actively locked FPCs has high complexity on both the optical and electrical system. They must be tuned and aligned every time the system turns on. Particularly when the cascaded FPCs are used, alignment and locking procedures can be much more lengthy and difficult.

High fundamental repetition rate ($f_{rep}$) is vital to obtain high side-mode suppression. Employing Ti:sapphire laser as the source comb helps to achieve GHz level fundamental repetition rate. However, the spectrum is not well repeated every time the Ti:sapphire laser initiates mode locking. They are also very bulky and costly. Fiber lasers can perfectly remedy these drawbacks, but currently fiber based astro-combs all suffer from low fundamental repetition rates, which need cascaded FPCs.

The order of super-continuum generation and mode spacing multiplication is another critical issue. If super-continuum is before the FPCs, it will be very difficult for the mirror coatings to meet both high reflection and low group delay dispersion (GDD) over an ultra-broad bandwidth. Cascaded FPCs will be unavoidable. Nevertheless, a bigger trouble lies there if the mode space

a)E-mail: zhgzhang@pku.edu.cn

multiplication is in the front, that tens of watts average power will be necessary for the super-continuum generation by using tapered photonic crystal fiber (PCF). Such high average power can easily damage the thin PCF in not long time. Amplification process can also asymmetrically re-amplify the filtered side-modes [11].

To solve the above issues, we propose a novel and compact astro-comb, which employs a simple scheme of "GHz fiber laser → spectrum broadening → ultra-low expansion (ULE) glass spaced FPC". In a previous conference paper [12], we have put forward the preliminary design. In this paper, we report details and improvements of this astro-comb, particularly its good qualities and feasibility.

## EXPERIMENTAL SETUP

Figure 1 shows the whole system architecture, including the optical part and the servo loops. Details will be discussed in the following.

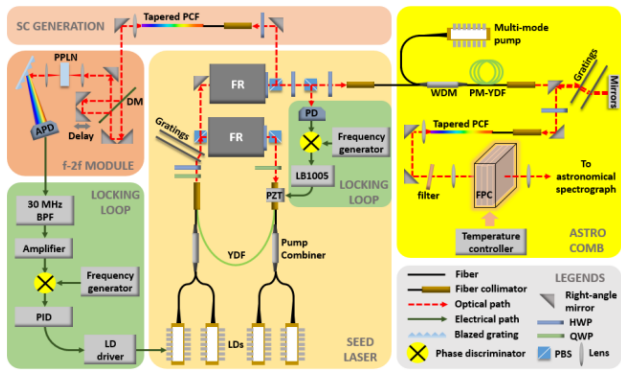

FIG. 1. The configuration of entire astro-comb. LD: laser diode; YDF: ytterbium doped fiber; HWP: half-wave plate; QWP: quarter-wave plate; PBS: polarization beam splitter; FR: Faraday rotator; DM: dichroic mirror; PPLN: periodically poled lithium niobate crystal; APD: avalanche photo detector; WDM: wavelength division multiplex; PM-YDF: polarization maintaining double cladding ytterbium doped fiber; BPF: band pass filter; LB1005: New Focus LB1005 servo controller.

### The seed comb

First, we chose the home-made high repetition rate (~837 MHz) mode locked Yb:fiber laser [13] as the seed laser to guarantee high side-mode suppression rate, good power scalability, reproducibility, and compactness at the same time. Although 1 GHz or higher $f_{rep}$ is preferred, we sacrificed a bit $f_{rep}$ for a better spectrum broadening and more dynamic range of LD current. Benefited from the high power (>600 mW) and intrinsically short pulse duration (<100 fs) of the seed laser, the output pulses could directly generate octave spanning super-continuum (SC) spectrum by a piece of tapered PCF without amplifier and compressor [14]. We obtained the carrier envelop frequency ($f_{ceo}$) with 40 dB signal to noise ratio (SNR) under 300 kHz resolution bandwidth (RBW) by a conventional f-to-2f method. The $f_{ceo}$ (30 MHz) and $f_{rep}$ (~837 MHz) were simultaneously locked to frequency synthesizers synchronized to a Rb clock by feedback to pump current and intra-cavity piezoelectric transducer, respectively. The frequency instability of the fully stabilized comb was ~$2\times10^{-12}$ at 1 s.

### The fiber amplifier and the spectrum broadening

Surplus power beyond $f_{ceo}$ and $f_{rep}$ detection was amplified to 2 W by a polarization maintaining double-cladding fiber amplifier. After compressor, the ~100 fs pulses were coupled into another piece of tapered PCF for SC generation of visible spectrum (560 nm to 700 nm). Such low average power never burns the PCF whose waist core diameter was ~2 micron. We spliced the PCF to a commercial compact single mode fiber (SMF) pigtailed collimator with insertion loss below 0.5 dB. The mismatch of mode field diameters between PCF and SMF could be solved by asymmetric discharging [15]. The collimators were integrated and designed in pair, pledging high (>80%) and stable coupling efficiency of the PCF. This coupling mode helped to hold the SC spectrum as stable as possible and saved space meanwhile.

### The Fabry-Perot cavity

After a short pass filter, the visible part of the SC spectrum was then sent into a home-made FPC for multiplication of mode spacing. To eliminate the active dithering of the locked FPC, we proposed a novel, passively stabilized FPC made by ULE glasses [16], as shown in Fig. 2. Two mirrors and a hollow spacer were assembled by optical contact, and the cavity length was determined by the thickness of the spacer. We put this FPC inside a small sealed chamber and controlled the temperature within ±0.05 K.

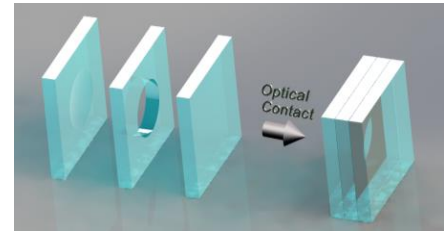

FIG. 2. Diagram of the hollow ULE glass spaced FPC.

The FPC was basically a plano-concave cavity. The outlines of the mirrors and the spacer were generally 40 mm × 40 mm × 5 mm sized but with different detailed structures. The plane mirror had a wedge angle of 30' to prevent from parasitic etalon effect, and the other mirror had a spherical area (200 mm radius of curvature) in the middle serving as the concave mirror. To ensure all the desired comb lines correctly pass, we designed a pair of complementary chirped multilayer coatings and coated them on the inward surfaces of the two mirrors. Anti-reflective coatings were coated on the outward surfaces. The spacer was a piece of 40 mm × 40 mm × 4.9 mm parallel ULE glass with a 20 mm diameter through-hole in the middle and a small through-hole on one side to balance the air pressure. The free spectral range ($\Delta v_{FSR}$) of this length-fixed FPC was thus a nonadjustable value about 29.3 GHz. To match this nonadjustable FPC, we searched for the best value of $f_{rep}$ by monitoring the filtered spectrum while changing the $f_{rep}$. When there was no "Vernier-like" pattern [1] and all the desired modes transmitted the most, the best $f_{rep}$ value was found to be 837.412090 MHz. The $\Delta v_{FSR}$ was then determined as 29.309423150 GHz (35 × $f_{rep}$).

### The mirror coatings

The reflection and dispersion of the mirror pair play the vital role during the filter process. We carefully designed a complementary chirped mirror pair with 99% typical round-trip reflection and constant round-trip group delay (GD) from 440 nm to 700 nm. The theoretical nonlinear round-trip phase shift was <3 mrad within this bandwidth. The designs of the two film structures are shown in Fig. 3(a), and their simulated properties are shown in Fig. 3(b). Ultra-thin (<10 nm) layers were removed so that the coatings were robust to the deposition conditions and insensitive to the manufacturing errors during the actual coating process [17]. For probing the possible errors, we simulated a random manufacturing error of 1% at most for each layer's thickness and performed 50 tests to estimate the possible phase errors [18]. The simulation result in Fig. 3(c) shows that possible round-trip phase error would spread within a maximum of 25 mrad which was acceptable.

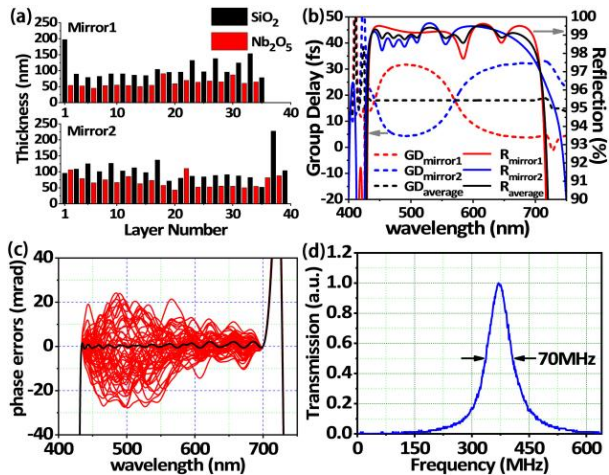

FIG. 3. (a) Film structures of the chirped mirror pair. (b) Simulated GD curves and reflection curves of the two mirrors and the round-trip average. The average curves are calculated by the following equations: $GD_{average} = (GD_{mirror1} + GD_{mirror2}) / 2$, and $R_{average} = \sqrt{R_{mirror1} \times R_{mirror2}}$. (c) Simulated nonlinear round-trip phase error of the original designs (black curve) and 50 tests (red curves) assuming a random deposition error of 1% for each layer's thickness. (d) Single transmission peak of the FPC. The full span corresponds to the 640 MHz sweeping range.

The finesse was measured by sweeping a narrow linewidth CW laser at 679 nm and measuring its transmission curve (Fig. 3(d)). The sweeping span was 640 MHz, then the full width at half maximum (FWHM) of the transmission was calculated to be ~70 MHz, corresponding to a typical finesse of about 420. Another CW laser at 461 nm was also employed to do the same experiment, obtaining a similar FWHM of ~85 MHz (finesse ~340). Those results support the agreement between the experiment implementation and the simulation. According to the measured finesse and the $f_{rep}$, the side-mode suppression ratio could be calculated to be >25 dB.

Finally, the filtered comb was coupled into a piece of multi-mode fiber with fiber shaker and mode scrambler, and then sent into the astronomical spectrograph (49,000 resolution). The inbuilt CCD camera imaged the two-dimension (2D) spectrum by exposure procedure.

## EXPERIMENT RESULTS AND DISCUSSIONS

### Exposure data and fittings

The exposure intensity data read from the CCD pixel array is plotted in Fig. 4(a). It indicates that at least 16 echelle orders were covered by high exposure intensity, and the total amount of highly exposed comb lines could be estimated as ~3,000. Figure 4(b) shows part of the real shot of the CCD imaging, from which one can see the overall uniformity. The SC spectrum didn't fill the whole bandwidth of the mirror coatings, hence more calibration lines (>8,000) are possible by a better SC generation.

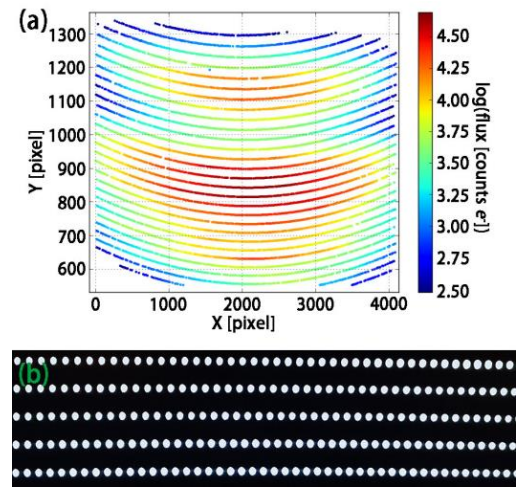

FIG. 4. (a) 2D spectrum data read from the CCD of astronomical spectrograph. (b) Photo of the CCD imaging. Each white dot stands for an individual comb line.

To have a more quantitative analysis, we picked out a single row data of one echelle order, fitted and plotted them in Figs. 6(a)-6(c). The original data were highly coincident to the Gaussian fitting curves except for a small mutual trail on the bottom of each mode. This trail was brought by a slight off-axis between fiber and lens inside the spectrograph, not the comb. It was confirmed by another lamp exposure and could be amended by post data processing. Considering the resolution was 49,000 and ~2.2 GHz occupied 1 pixel, all the exposed comb lines reached the resolution-limited FWHM of ~4.5 pixels. Despite the very high exposure intensities, the noise floor between two neighbor comb lines still remained zero, which means that all neighbor comb teeth could be clearly separated. The zero noise floor allowed the highest exposure SNR to be >200. On the other hand, the peak intensities of the comb lines were close to each other. Such good uniformity made all mode able to be highly and equally exposed. These phenomena indicated that the desired comb lines were all correctly filtered. The good linetype, sharp profile, zero noise floor and uniform peaks were greatly benefited from the "quiet (no dithering)" character of the ULE cavity. By fitting the centers of gravity, the optical frequencies of comb modes showed a linear correlation with their series number with a slope of K=29.31 GHz (see Fig. 5(d)).

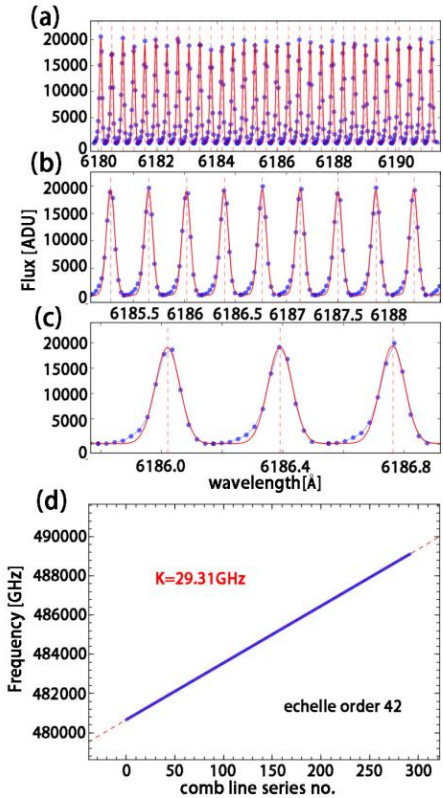

FIG. 5. (a)-(c) Exposure results (blue dots and connections) and their Gaussian fittings (red) under different scales. (d) Line space fitting of the comb lines. K is the slope.

We could computed the RV precision according to the formula [19],

$$\sigma_v \approx \delta v / \sqrt{N} \approx A \frac{\text{FWHM}}{\text{SNR} \times \sqrt{n}} \times \frac{1}{\sqrt{N}} \quad (1)$$

Where A is a constant depend on the functional form of the line profile, FWHM is the frequency width of a single comb line in term of m/s, SNR the signal to noise ratio of the comb lines, $n$ the number of pixels occupied by a single comb line, $N$ the total number of comb lines included for calibration. The typical values of the parameters are: A = 0.41 [19], $n$ = 4 and FWHM = 5.6 km/s for our spectrograph. The SNR of the exposure result was conservatively averaged to be 130. By taking clearly resolved 3,000 comb lines (i.e. $N$ = 3,000), we computed the RV precision was ~16 cm/s.

## Discussions

The FPC spaced by a hollow ULE glass has an excellent short term stability which is favorable to improve the quality of the exposed comb lines. On the other hand, however, long term drift of the FPC would asymmetrically leak the side-mode over time and shift the centers of gravity to some extent. Although we controlled the temperature carefully, the slow drift could still be observed strongly correlated with the air pressure over days. However, this issue could be overcame by employing a very small vacuum chamber. In that case, the cavity length will be simply decided by the temperature. According to the datasheet of the ULE glass [16], the maximum thermal expansion coefficient of the ULE glass under room temperature is $3 \times 10^{-8}$ /K. Considering the 0.1 K temperature variation, >25 dB side-mode suppression and 837 MHz $f_{\text{rep}}$, the maximum drift of the cavity length could only cause a fitting center drift for 5 kHz. The influence on the RV measurement will be (5 kHz/500 THz) $\times 3 \times 10^{10}$ cm/s = 0.3 cm/s, which is very small. A 187 mm long ULE glass FPC was measured in the vacuum, proving $10^{-9}$ level long term length change [20]. Since our cavity is much shorter, other factors like mechanical deformation or aging will be much smaller, so the ULE glass FPC is very promising for astro-combs.

Such a length-fixed FPC greatly simplified the whole system, as the CW laser and PDH locking loop are no longer need. Due to the high repetition rate of the seed fiber laser and the broadband chirped mirror coatings, cascaded FPCs are also unnecessary. Including the collimator spliced PCF, all the procedures made the astro-comb compact, stable, easily built, and cost efficiency while improving the performance. It's definitely of great advantages to astro-combs, which are calibration instruments in nature.

## CONCLUSION

We demonstrated a compact and robust 29.3 GHz astro-comb based on a 837 MHz Yb:fiber laser. The GHz level fiber laser comb stabilized to Rb clock laid the foundation of high stability and high side-mode suppression. A passively stabilized FPC made by ULE glass is innovatively applied in the astro-comb for the first time. The comb offered ~3,000 calibration lines from 560 nm to 700 nm. Benefited from the good short term stability of ULE glass spacer and the well-designed mirror coatings, the exposed calibration lines had very sharp linetype, high contrast, and good uniformity. Limited by the bandwidth of the SC and uneven spectral intensity, the RV precision was computed to be ~16 cm/s. Extending the spectrum to 400 nm could increase the comb lines to above 8,000 and promote the calibration resolution to several cm/s level. This comb was simple, compact, robust, easily operated and cost effective. In addition, it can easily adapt to other astronomical spectrographs with different resolution by simply replacing the spacer of the FPC, showing a good compatibility.

## FUNDING

## REFERENCES AND FOOTNOTES